\documentstyle[aps,prl,preprint,floats,epsfig]{revtex}

\begin{document}

\title{ \vspace{-2cm} \quad\\[1cm] \Large \setlength{\baselineskip}{20pt}
Measurement of the inclusive semileptonic branching \\
fraction
   of
{\it \boldmath B} mesons  and $|V_{cb}|$}

\maketitle
\tighten

{\vspace{-3cm}\noindent
{\hspace{10cm} BELLE Preprint 2002-26}\\
{\hspace{10cm} KEK Preprint 2002-81}
}

\vspace{2.5cm}
\begin{center}
{\large (The Belle Collaboration)}
\vspace{0.5mm}

   K.~Abe$^{9}$,               
   K.~Abe$^{44}$,              
   T.~Abe$^{45}$,              
   I.~Adachi$^{9}$,            
   Byoung~Sup~Ahn$^{16}$,      
   H.~Aihara$^{46}$,           
   M.~Akatsu$^{23}$,           
   Y.~Asano$^{51}$,            
   T.~Aso$^{50}$,              
   V.~Aulchenko$^{2}$
% ...
\end{center}

\small
\begin{center}
% ...
$^{2}${Budker Institute of Nuclear Physics, Novosibirsk}\\
% ...
$^{9}${High Energy Accelerator Research Organization (KEK), Tsukuba}\\
% ...
$^{16}${Korea University, Seoul}\\
% ...
$^{23}${Nagoya University, Nagoya}\\
% ...
$^{44}${Tohoku Gakuin University, Tagajo}\\
$^{45}${Tohoku University, Sendai}\\
$^{46}${University of Tokyo, Tokyo}\\
% ...
$^{50}${Toyama National College of Maritime Technology, Toyama}\\
$^{51}${University of Tsukuba, Tsukuba}\\
% ...
\end{center}

\normalsize


\begin{abstract}
We present a measurement of the electron spectrum from inclusive
semileptonic {\it B} decay,
using 5.1~fb$^{-1}$ of $\Upsilon(4S)$ data collected
with the Belle detector. A high-momentum lepton tag was used to
separate the semileptonic {\it B} decay electrons
from secondary decay electrons.
We obtained the branching fraction,
${\cal B}(B\rightarrow X e^+ \nu) = (10.90 \pm 0.12 \pm 0.49)\%$,
with minimal model dependence.
   From this measurement, we derive a value for the 
Cabibbo-Kobayashi-Maskawa matrix element 
$|V_{cb}| = 0.0408 \pm 0.0010 {\rm (exp)} \pm 0.0025{\rm (th)}$.

\end{abstract}

\vspace{1cm}



{\renewcommand{\thefootnote}{\fnsymbol{footnote}}

\normalsize
\setcounter{footnote}{0}
\newpage

\normalsize

%
%
         \section{ Introduction }
%
The inclusive semileptonic branching fraction of {\it B} decay
has long been an interesting puzzle in heavy flavor physics.
Bigi et al.~\cite{puzzle} first pointed out that
theoretical calculations including QCD corrections
disagreed with experimental results.
While most measurements have consistently been smaller than
11\%~\cite{{expA-br},{expC-br},{expZ-br}},
theoretical expectations have been higher than 12\%.
Some theoretical analyses have been better able to
accommodate a low
semileptonic branching fraction
by including final state mass effects in the next-to-leading order
QCD corrections ~\cite{{th-ccs},{b2ccs}}.
This does not necessarily solve the puzzle, however, because the
semileptonic branching fraction is correlated with the rate for
${\cal B}( b\rightarrow c \bar{c} s )$, which depends on both the
quark-mass ratio ($m_c/m_b$) and  the renormalization scale
($\mu/m_b$), where
$\mu$ is used to renormalize the coupling
constant ($\alpha_s(\mu)$) and Wilson coefficients ($c_{\pm}(\mu)$)
appearing in the non-leptonic decay rate.
Current measurements of the charm multiplicity in $B$ decay
($n_c$)\cite{Nc_CLEO} cannot easily accommodate the semileptonic
branching fraction measured at the $\Upsilon$(4S) resonance,
although measurements at higher energies~\cite{{Nc_DELPHI},{Nc_ALEPH}}
are somewhat more compatible.
Further studies at the $\Upsilon$(4S) are needed to achieve a
better understanding of $B$ decay and theoretical models.
In addition, the semileptonic branching fraction may be combined with
the decay lifetime to obtain the partial width, which is used
to extract the Cabibbo-Kobayashi-Maskawa (CKM)
matrix element $|V_{cb}|$~\cite{KM}
and to probe
theoretical models of {\it B} decay.

For this measurement, it is essential to distinguish
between primary decay leptons from
$B\rightarrow X\ell^+\nu$
and secondary decay leptons produced mainly through charm decay
($B\rightarrow \bar{D}X, \bar{D} \rightarrow Y\ell^-\bar{\nu}$).
We have used a dilepton method introduced by ARGUS~\cite{expA-br}
to minimize model dependence  in this measurement.
This approach requires
a high momentum lepton, an electron or a
muon, to identify a $B\bar B$ event and to tag the flavor of one of them.
We then study the spectra of additional leptons in the event,
extended to low momenta, selected so that they come mainly from
decays of the other $B$, and separated by  charge relative to the
tagging lepton 
to  distinguish primaries and secondaries effectively.
This spectrum study is performed using electrons only, 
as their experimental identification
extends to much lower momenta than that of muons.
The numbers of electrons in our sample are determined by fitting the
distributions of the ratio of cluster energy to track momentum
($E/p$) in each kinematic bin.
The branching fraction is then obtained by normalizing to the total
number of tag leptons, rather than the luminosity or the number of
$B\bar{B}$ events.

%
%
    In this paper, we report on the branching fraction of
inclusive semileptonic $B$ decay $(B \rightarrow X e^+ \nu)$
and the CKM matrix element $|V_{cb}|$.
Charge conjugation is implicitly included.
   The data sample used in this analysis was collected
at the $\Upsilon(4S)$ resonance
with the Belle detector~\cite{Belle} at the KEKB
asymmetric $e^+e^-$ (3.5 on 8~GeV) collider~\cite{KEKB}.
   This analysis is based on an integrated luminosity
of 5.1~fb$^{-1}$ at the $\Upsilon(4S)$ resonance,
and 0.6~fb$^{-1}$ continuum data at a
center-of-mass energy of 60~MeV below the $\Upsilon(4S)$.
A GEANT~\cite{MC} based Monte Carlo simulation
was used to model the detector response.

%
    \section{ The Belle detector }
%
The Belle detector is configured around a 1.5~T superconducting
solenoid and iron structure surrounding the KEKB interaction region.
It covers 92\% of the total solid angle
in the $\Upsilon(4S)$ center-of-mass (CM) system.
Charged-particle tracking is provided by
three layers of double-sided silicon vertex detectors~(SVD)
and a 50-layer central drift chamber~(CDC).
Eighteen of the wire layers provide a small stereo angle
to measure the coordinates of the particle trajectories in the
direction of the beam ($z$).
The transverse momentum resolution for charged tracks is
$ ( \sigma_{p_t}/p_t )^2 = ( 0.0019 p_t )^2 + ( 0.0030 )^2$,
where $p_t$ is in GeV/$c$.

Charged hadron identification is provided by
$dE/dx$ measurements in the CDC,
an aerogel \v{C}erenkov counter~(ACC),
and a time-of-flight scintillation counter (TOF).
The $dE/dx$ measurements have a resolution for
lepton tracks of 6\% and are useful for separating electrons
from hadrons over nearly the full momentum range.
The ACC and the TOF are used to reject charged $K$ and
proton tracks.

The electromagnetic calorimeter (ECL) contains 8736 CsI(Tl) crystals
located behind the hadron identification detectors, inside
the solenoid.
Its thickness is 16.2 radiation lengths over  the entire tracking acceptance.
The photon energy resolution is
$ ( \sigma_E/E )^2 = ( 0.013 )^2 + ( 0.0007 / E )^2
+ ( 0.008 / E^{1/4} )^2$, where $E$ is in GeV.

The $\mu/K_L$ detector (KLM) located outside the coil consists of
15 sensitive layers
and 14 iron layers in the octagonal barrel region.
The iron plates and associated materials provide a total of 4.7
nuclear interaction lengths at normal incidence.

%
    \section{ Dilepton Selection  }
%
Hadronic events were selected based on charged track information from
the CDC and cluster information from the ECL after transformation to
the CM system.
We required at least
three charged tracks, an energy sum in the calorimeter
between 10\% and 80\% of $\sqrt{s}$,
a total energy sum of greater than 20\% of $\sqrt{s}$, and
a total momentum balanced in the $z$ direction within
2.1~GeV/$c$.
This removed the majority of two-photon, radiative Bhabha, and
$\tau^+\tau^-$ events where both $\tau$'s decay to leptons.
Remaining radiative Bhabha events were removed by requiring at least two
large angle clusters in the ECL and that the average cluster energy be
below 1~GeV.
In order to remove higher multiplicity $\tau^+\tau^-$ events, we
calculated the invariant mass of the particles found in each hemisphere
perpendicular to the event thrust axis and removed the event if it fell
below the $\tau$ mass.
Beam-gas and beam-wall backgrounds were removed by reconstructing the
primary  vertex of the event and requiring it to be consistent with
the known location of the interaction point(IP).
To suppress continuum, we required that the ratio $R_2$ of second to
zeroth Fox-Wolfram moments~\cite{FW} be less than 0.6.


Events containing a high-momentum ``tagging lepton'' to tag the {\it
B} flavor and
a ``spectrum electron'' for the spectrum study were selected from
the hadronic event sample.
To select tagging electrons, we used  $dE/dx$
measurements in the CDC, the response of the ACC,
matching between the ECL cluster and associated CDC track,
the shower shape of the  ECL cluster,
and $E/p$, the ratio of the ECL energy to the
momentum of the associated CDC track.
Each set of measurements was translated into an electron probability.
These probabilities were then combined into a single likelihood discriminant,
$L_{e}$.
For the selection of tagging muons, we defined an equivalent likelihood
discriminant $L_{\mu}$
from information on the location and penetration depth of the associated
track in the KLM.
In addition to the requirements   $L_e > 0.8$ and $L_{\mu} > 0.95$,
the tagging lepton tracks were required to satisfy
1.4~GeV/$c<p<$2.2~GeV/$c$,
$45^{\circ}<\theta_{\rm lab} <125^{\circ}$,
$|dr_{IP}| <$ 0.2~cm, and $|dz_{IP}| <$ 10~cm, where $p$ is the CM
momentum, $\theta_{\rm lab}$ is the polar angle of the track in the
laboratory frame, and $dz_{IP}$ and $dr_{IP}$ are
the distances of closest approach to the IP in the direction of the
beam and in the plane perpendicular to it, respectively.
For  tagging electrons that satisfy the above selection criteria,
the detection efficiency, determined by embedding single simulated
tracks in multi-hadron data, is 92\%, and the misidentification
probability for hadrons,
determined from  $K^0_S \rightarrow \pi^+\pi^-$ decays, is less than 0.4\%.
For tagging muons, the efficiency and misidentification probability,
determined in a similar way, are 86\% and 1.2\%, respectively.

To select electron candidates for the spectrum measurement,
we required that  tracks  satisfy
$p_{\rm lab} > 0.5$~GeV/$c$, $ 46^{\circ}<\theta_{\rm lab}<125^{\circ} $
, $|dr_{IP}|<0.3$~cm and $ |dz_{IP}| < 4.0$~cm
($p_{\rm lab}$ is the momentum in the laboratory frame).
We used  $dE/dx$, TOF, and ACC information
to reject  tracks that are strongly identified as being other than electrons
by a likelihood criterion of $L_{non-e} < 0.95$.
The efficiency of this cut for retaining electrons was 98.6\%.
We did not use  ECL cluster information because that was used at a
later stage to
extract electron yields by fitting the $E/p$ distribution.


Further requirements were made to reduce backgrounds from several
known sources.
Events were rejected if any track satisfying the lepton
identification requirement of $L_e >0.8$ or $L_\mu > 0.95$ could be
paired with any other oppositely charged track to obtain an
invariant mass  within 50~MeV/$c^2$ of  the $J/\psi$ mass.
To remove electrons originating from
$\gamma$ conversions, we rejected all pairs of oppositely-charged
tracks where at least one track satisfied $L_e > 0.8$ and the 
$e^+e^-$ invariant mass was less than 0.10~GeV/$c^2$.
To remove electrons from $\pi^0$ Dalitz decays, oppositely charged
track pairs with $e^+e^-$ invariant mass less than 0.10~GeV/$c^2$
were combined with photon candidates and rejected if the
$e^+e^-\gamma$ invariant mass of any combination was  consistent with
the $\pi^0$ mass.

To reduce dileptons in continuum events, which tend to be
back-to-back, and ``ghost'' tracks (one track mis-reconstructed as
two nearly
collinear tracks),
we required $ -0.8 < {\rm cos}\theta_{\ell e} < 0.998$
for both opposite-sign and same-sign dileptons, where $\theta_{\ell e}$ is
the opening angle between the tagging lepton and the spectrum electron in
the CM frame.

%
%

To reject same-$B$ backgrounds, where both leptons originate from
just one $B$ meson, we look at angular, kinematic, and charge
correlations.
The spectrum electron and the tagging lepton, if correctly selected, have
opposite charges unless one of the parent $B$ mesons has mixed.
If a secondary electron is selected as the
spectrum electron, the charge combination is opposite to the above case.
If a secondary electron from the same $B$ as the tagging lepton is
selected as the spectrum electron, it has a charge opposite to that
of the tagging
lepton.
Opposite-sign dileptons were required to satisfy either
$ p_e + {\rm \cos}\theta_{\ell e} > 1.2$ ($p_e$ in GeV/$c$)
or $ {\rm \cos}\theta_{\ell e} > 0.3 $.
This criterion reduced the fraction of the same-$B$ background in the spectrum
electron candidates to 2\% while retaining 61\% of the signal.

%
    \section{ Signal yield and Background subtraction }
    \label{sec:sigyield}
    To extract the number of dilepton events from the lepton-tagged events,
   a maximum likelihood
fit was performed on the $E/p$ distributions of the
spectrum electron candidates in each momentum ($p$)
and  polar angle ($\theta_{\rm lab}$) bin
as shown in Fig.~\ref{eoverpfit},
where the momentum range is divided into 46 bins of width  0.05 GeV/$c$
(0.5--2.8 GeV/$c$),
the polar angle region is divided into 4 bins
(46--60$^\circ$, 60--78$^\circ$, 78--97$^\circ$ and
97--125$^\circ$), and same- and opposite-sign dileptons are
treated separately.
The signal probability density function (PDF) was obtained using
$e^+$ and $e^-$ from photon conversions and 
from the two-photon process.
The background PDF was taken from  the sample of
tracks rejected by the
electron identification requirements for the spectrum study.
By determining both signal and background PDF's in the same bins of
momentum and angle as the dilepton sample,
we removed any systematic effects due
to differences in momentum and angular distribution between the
samples used to obtain the PDF's and our dilepton sample.

The raw momentum spectra
in the CM frame for both opposite-sign
and same-sign electrons with the above selections applied are shown
in Fig.~\ref{spectrum1}.
The total yields were
$19722\pm 147$ opposite-sign (OS) and $11224\pm 117$ same-sign (SS) 
dileptons in the data taken at
the $\Upsilon(4S)$, and $791\pm 105$ OS and $448\pm 87$ SS  below resonance
   (after scaling).
The raw yields are shown in Table~\ref{e_yield}.
The continuum contribution was subtracted by scaling the dilepton
yields in the off-resonance data.
The scaling factor is the ratio of the on- and off-resonance
integrated luminosity,
after correcting for the energy dependence
of the continuum cross section.

%
%

These uncorrected spectra contain several backgrounds that are
accounted for and subtracted.
Leptons from $J/\psi$ decay and electrons from $\gamma$ conversion
and $\pi^0$ Dalitz decays can survive
when one of the pair has escaped detection.
These backgrounds are determined by Monte Carlo simulation, and
the uncertainties were evaluated from the error on each rate.
Contributions from previously unaccounted sources of secondary
leptons from $B$ decay were modeled via Monte Carlo, assuming current
values of branching fractions:
$B\rightarrow X \tau^+ \nu, \tau^+\rightarrow e^+Y$ (${\cal
B}(B\rightarrow X \tau^+ \nu)=(2.6\pm 0.4)\%$~\cite{PDG2000}),
$B\rightarrow D_sX, D_s\rightarrow e^-Y$ (${\cal B}(B\rightarrow
D_sX){\cal B}(D_s\rightarrow e^-Y) =(0.8\pm 0.3)\%$),
$B\rightarrow\Lambda_cX, \Lambda_c\rightarrow e^+Y$ (${\cal
B}(B\rightarrow \Lambda_cX){\cal B}(\Lambda_c\rightarrow
e^+Y)=(0.29\pm 0.12)\%$),
$B \rightarrow D X_{\bar{c}}, D \rightarrow Y e^+ \nu$ via
$\bar{b}\rightarrow\bar{c} c \bar{s}$
(${\cal B}(B\rightarrow D X_{\bar{c}})$
= (7.9 $\pm$ 2.2)\%~\cite{CLEO_DDX}).
%
False tag leptons arise mainly from secondary leptons and fake muons,
whose contamination was estimated to be
$(2.2\pm 0.4)\%$ and $(1.5\pm 0.8)\%$, respectively.
All backgrounds in the dilepton events
were scaled by normalizing to the number of lepton-tagged events.
The background yields and their errors are
summarized in Table~\ref{e_yield}.
The spectra of the total $B\bar{B}$ backgrounds
are included in Fig.~\ref{spectrum1}.

%
%
    \section{ Measurement of Branching fraction }
%
The opposite- and same-sign electron spectra after background
corrections can be written as
\begin{eqnarray}
   \frac{dN_{+-}}{dp} = \epsilon_{k1}(p) \eta(p) N_{\rm tag}
    \left[ \frac{d{\cal B}(B\rightarrow X e^+ \nu)}{dp} ( 1 - \chi )
     + \frac{d{\cal B}(B\rightarrow \bar{D}X, \bar{D} \rightarrow Y e^-
\bar{\nu})}{dp} \chi \right],
   \label{mixing1}             \\
   \frac{dN_{\pm\pm}}{dp} = \epsilon_{k2}(p) \eta(p) N_{\rm tag}
    \left[ \frac{d{\cal B}(B\rightarrow X e^+ \nu)}{dp} \chi
           + \frac{d{\cal B}(B\rightarrow \bar{D}X, \bar{D}
\rightarrow Y e^- \bar{\nu})}{dp}
                                                        ( 1 - \chi ) \right],
   \label{mixing2}
\end{eqnarray}
where $N_{\rm tag}$ is the number of tag leptons,
$\eta(p)$ is the electron identification efficiency as
a function of the momentum,
which also includes the  momentum smearing effect from bremsstrahlung, and
$\epsilon_{k}(p)$ is the efficiency of our kinematic selection, 
determined from a Monte Carlo simulation.
  $\chi$ is an effective  mixing parameter for all $B$ mesons coming from
$\Upsilon(4S)$ decay, $\chi \equiv \chi_0 f_{00} $, where $\chi_0$ is 
the fraction of neutral $B$ events decaying as mixed ($B^0B^0$ or 
$\bar B^0\bar B^0$) and $f_{00}$ ($f_{+-}$) is the branching fraction 
for $\Upsilon(4S)$
decay into $B^0\bar{B^0}$($B^+B^-$).
We used $\chi = 0.0853 \pm 0.0055$, calculated using
$\chi_0 = 0.174 \pm 0.009 $, $f_{+-}/f_{00} = 1.04 \pm 0.08$~\cite{PDG2000},
and $f_{+-} + f_{00} = 1.0 $.
The two spectra, $d{\cal B}(B\rightarrow X e^+\nu)/dp$ and
$d{\cal B}(B\rightarrow \bar{D}X,\bar{D} \rightarrow Y e^-\bar{\nu})/dp$,
can be obtained 
from the measured $dN_{+-}/dp$ and $dN_{\pm\pm}/dp$ spectra by
solving equations~(\ref{mixing1}) and (\ref{mixing2}) simultaneously.

%
%
The number of tags was determined by counting leptons above 1.4~GeV/$c$,
subtracting backgrounds, and correcting for the relative efficiency
for selecting dilepton events compared to single-lepton events
[(94.8$\pm$0.5)\%].
As a consequence, we found the number of tags to be
$ N_{\rm tag} = 528975 \pm 1412 ({\rm stat}) $.
The difference in the relative efficiency between opposite-sign and
same-sign dilepton events was also corrected for in
each electron spectrum.

%
%
By solving the equations, we obtained the momentum spectra for
both the primary and secondary electrons, as shown in Fig.~\ref{final}.
The primary electron spectrum was integrated from 0.6 to 2.8~GeV/$c$
to extract the partial branching fraction for semileptonic $B$ decay,
\begin{equation}
   {\cal B}( {\it B} \rightarrow Xe^+ \nu ~|~ p > 0.6~{\rm GeV/}c )
   = ( 10.24 \pm 0.11{\rm (stat)} \pm 0.46{\rm (syst)} )\%.
\end{equation}
%
%
The relative size of the  undetected region of
the electron spectrum was studied
using the ACCMM~\cite{ACCMM} and ISGW2~\cite{ISGW} models
and a semi-empirical shape~\cite{QQ} for $b\rightarrow c$ decay,
with options to include semileptonic decay with baryons~\cite{QQ},
and charmless semileptonic $B$ decays.
As part of our evaluation, we fitted the measured primary spectrum to
the prediction of each model.
Results of the fitting are shown in Table~\ref{tab:chisq}.
For all models, the fits were improved when the proportion of decays
$B\rightarrow D^{**}\ell\nu$ was allowed to float to higher values
than were nominally set in the model.
We obtained the best $\chi^2$ fit using a semi-empirical shape, with
${\cal B}({\it B}\rightarrow  X_u e\nu)$ fixed to 0.167\%, and the
following quantities allowed to vary (fitted values with statistical
error are shown):
${\cal B}({\it B}\rightarrow  D^{(*)}e\nu) = 7.31 \pm 0.28$\%,
${\cal B}({\it B}\rightarrow  D^{**}e\nu) = 2.51 \pm 1.00$\%,
${\cal B}({\it B}\rightarrow  D^{(*)}\pi e\nu) = 0.78 \pm 0.96$\% and
${\cal B}({\it B}\rightarrow  {\rm baryons}~ e\nu) = 0.03 \pm 0.31$\%.
We extrapolated the branching fraction below 0.6 GeV/c using the
electron momentum spectrum of this ``best-fit'' model.
We found that
the extrapolated area was  $(6.1\pm0.7)\%$ of the total decay
width,
where the error was taken from the largest difference between
the best-fit model and the other models (excepting ACCMM, which gave
a poor fit).
We obtained:
\begin{equation}
   {\cal B}( {\it B} \rightarrow Xe^+ \nu )
   = (10.90 \pm 0.12{\rm (stat)} \pm 0.49{\rm (syst)})\%.
\label{final_value}
\end{equation}
This result is consistent with other
measurements~\cite{{expC-br},{expZ-br}}.

%
To determine the CKM matrix element $|V_{cb}|$ we first subtracted
the contribution from charmless decay (${\cal B}(B\rightarrow X_u
\ell \nu)$=(0.167$\pm$0.055)\%)  and assumed ${\cal B}(B\rightarrow X
e^+ \nu) = {\cal B}(B\rightarrow X \ell\nu)$.
Using a formula based on the heavy quark expansion~\cite{Bigi},
with $\mu_{\pi}^2=(0.5\pm 0.2)\ {\rm GeV}^2$, $m_b =4.58\pm 0.05\
{\rm GeV}/c^2$
and the world-average value for $B$ lifetime, $\tau_B$ = 1.607$\pm$0.021~ps
(average of  $B^0$ and $B^+$)~\cite{PDG2000},
our evaluation is
\begin{equation}
    |V_{cb}| = 0.0408 \pm 0.0010{\rm (exp)} \pm 0.0025{\rm (th)},
\end{equation}
where the first error includes both the statistical and systematic
uncertainties and the second error arises from theoretical
uncertainties.

The branching fraction for secondary decay,
${\cal B}(B\rightarrow \bar{D}X ,\bar{D}\rightarrow  Y e^-
\bar{\nu})$, was extracted by fitting the secondary electron spectrum
to the ACCMM and ISGW2 model predictions.
We found
\begin{eqnarray}
{\cal B}(B \rightarrow \bar{D}X,\bar D \rightarrow Y e^- \bar{\nu})
= ( 7.06 \pm 0.14{\rm (stat)} \pm 0.48{\rm (syst)} )\%\ (ACCMM)\\ \nonumber
= ( 7.84\pm 0.13{\rm (stat)} \pm 0.48{\rm (syst)} )\%\ (ISGW2)
\end{eqnarray}
The obtained branching fraction of the secondary charm semileptonic
decay is consistent with
results from other experiments~\cite{{CLEO_DDX},{PDG2000}}.

%
%
    \section{ Systematic errors }
%
The systematic uncertainties on
${\cal B}(B\rightarrow X e^+ \nu)$ and
${\cal B}(B\rightarrow \bar{D} X ,\bar{D}\rightarrow Y e^- \bar{\nu})$
are summarized in Table~\ref{sys_table}.
The main contributions are the uncertainties in
the tracking and identification efficiencies for the electrons used
in the spectrum study.
%
%
The tracking efficiency was estimated  by embedding  electron
tracks  into hadronic event data
and repeating the
tracking procedures to evaluate the efficiency of the
embedded electron.
The  embedded  electrons were obtained  from
single electron Monte Carlo as well as
from electrons in two-photon data.
The tracking efficiency was also estimated by embedding
into simulated $B\bar{B}$ events, and the absolute
difference in values from the two methods was taken as the
uncertainty.

The electron identification included two steps: hadron-muon rejection
and $E/p$ fitting.
The inefficiency in the first step (1.4\%) was found from a study
using photon conversion electrons embedded into hadronic events in
data.
The systematic uncertainty of 0.8\% on this effect  was estimated by taking the
difference between results obtained from embedding in data and MC.
Uncertainties in $E/p$ fitting can come from impurities in the data
samples used to obtain the signal and background shapes as well as
differences in the event environment between those data samples and
dileptons.
The signal shapes were obtained from
photon conversion electrons in the hadronic sample and two photon data,
while the background shapes were obtained from tracks failing the
hadron-muon rejection.
Impurities in these samples were estimated via Monte Carlo, and
systematic uncertainties of 0.9\% and 0.8\% were assigned for fitting
of electron and hadron shapes, respectively.
The uncertainty due to differences in event environment was also
estimated using MC and found to be 1.3\%.
The quadratic sum of the contributions gives a net uncertainty on
electron identification of 2.0\%.

The uncertainty in the kinematic selection efficiency was estimated
by changing the event selection criteria for signal events.
Instead of requiring  either $p_e + \cos\theta_{\ell e} > 1.2$
or $\cos\theta_{\ell e} > 0.3$,  we
required either $p_e + \cos\theta_{\ell e} > 1.0$ or
$\cos\theta_{\ell e} > 0$.
We also varied
the polar angle region for tag leptons and considered the ranges;
$45^\circ < \theta_{\rm lab} < 65^\circ$,
$ 65^\circ < \theta_{\rm lab} < 90^\circ$ and
$90^\circ < \theta_{\rm lab} < 125^\circ$.
The maximum deviation of the results was assigned
to a systematic error.

The systematic uncertainty in the continuum subtraction was
attributed to the uncertainty on the normalization between
on-resonance and off-resonance.
The uncertainty was estimated from the error on  the integrated
luminosity.
This effect  is larger
for the lepton tag events than the dilepton events since the fraction of
continuum background for tag events is much higher than for dilepton events.

There were two contributions to the uncertainty on the background
subtraction in the dilepton events.
The first is the uncertainties on the branching
fractions of the various sources, which are given in
Section~\ref{sec:sigyield}.
The second is the uncertainty on the scale factor
used to normalize  background yields obtained  from Monte Carlo simulation.
The dominant contributor to this effect is the uncertainty in the 
number of tag leptons, which was used to determine scaling factors.

The main sources for the uncertainty on the relative efficiency are
the invariant mass cuts to reject electrons from $J/\psi$ and gamma 
conversions.
This uncertainty was estimated by taking the difference of the final yields
with and without the cuts.

%
%
     \section{ Conclusion }
%
In conclusion,
we have measured the branching fraction for inclusive semileptonic
$B$ decay using a high-momentum lepton tag.
The result is
\begin{equation}
{\cal B}( B \rightarrow X e^+ \nu ) = (10.90 \pm 0.12 \pm 0.49)\%,
\end{equation}
where the first error is statistical and the second is systematic.
This result agrees with other measurements.
  From this result and the world average $B$ meson lifetime~\cite{PDG2000},
we have extracted the CKM parameter $|V_{cb}|$ as follows:
\begin{equation}
    |V_{cb}| = 0.0408 \pm 0.0010{\rm (exp)} \pm 0.0025{\rm (th)}.
\end{equation}

\vspace{10mm}

\begin{acknowledgements}

We wish to thank the KEKB accelerator group for the excellent
operation of the KEKB accelerator.
We acknowledge support from the Ministry of Education,
Culture, Sports, Science, and Technology of Japan
and the Japan Society for the Promotion of Science;
the Australian Research Council
and the Australian Department of Industry, Science and Resources;
the National Science Foundation of China under contract No.~10175071;
the Department of Science and Technology of India;
the BK21 program of the Ministry of Education of Korea
and the CHEP SRC program of the Korea Science and Engineering Foundation
and the Center for High Energy Physics sponsored by the KOSEF
and the Yonsei University Faculty Research Support program;
the Polish State Committee for Scientific Research
under contract No.~2P03B 17017;
the Ministry of Science and Technology of the Russian Federation;
the Ministry of Education, Science and Sport of the Republic of Slovenia;
the National Science Council and the Ministry of Education of Taiwan;
and the U.S.\ Department of Energy.

\end{acknowledgements}


\newpage

%
%

\begin{table}[htbp]
\caption{ Yield of electrons with  momentum
of 0.6~GeV/$c<p_e<$2.8~GeV/$c$ in lepton-tagged events.
The background yields and their statistical errors were
determined by a Monte Carlo simulation.
The second errors are due to branching fraction uncertainties.  }
\label{e_yield}
\begin{tabular}{lrr}
                  & Opposite sign      & Same sign               \\
                              \hline \hline
On-resonance data & 19722 $\pm$ 147   & 11224 $\pm$ 117   \\
Scaled off-resonance
                    &   791 $\pm$ 105   &   448 $\pm$ \thinspace~87   \\
Continuum subtracted
                    & 18931 $\pm$ 181   & 10776 $\pm$ 146   \\  \hline
Lepton from J/$\psi$ or $\psi$'
                 &   167 $\pm$  ~8 $\pm$ ~33 & 179 $\pm$  ~8 $\pm$ ~36 \\
$e$ from $\gamma$ &   127 $\pm$  ~7 $\pm$ ~32 & 306 
$\pm$\negthinspace~10 $\pm$ ~77 \\
$e$ from $\pi^0$, $\eta$
                 &    30 $\pm$  ~3 $\pm$ \thinspace~~8 &  88 $\pm$  ~6 
$\pm$ ~22 \\
$e$ from $\tau$  &   406 $\pm$\negthinspace~12 $\pm$ ~84 &  75 $\pm$ 
~5 $\pm$ ~16 \\
$e$ from $\Lambda_c$
                 &     8 $\pm$  ~2 $\pm$  \thinspace~~3 & 177 $\pm$ 
~8 $\pm$ ~71 \\
$e$ from $D_s$   &   293 $\pm$ 10 $\pm$ ~88 &  64 $\pm$  ~5 $\pm$ ~19 \\
$e$ from $D (B \rightarrow D X)$
                 &   285 $\pm$\negthinspace~10 $\pm$ ~86 &  54 $\pm$ 
~4 $\pm$ ~16 \\
Same-$B$       &   397 $\pm$\negthinspace~12 $\pm$ ~79 &        ---        \\
Tag from {\it D} & 160 $\pm$  ~8 $\pm$ ~32 & 600 
$\pm$\negthinspace~15 $\pm$ 120 \\
Fake lepton tags &   240 $\pm$  ~9 $\pm$\negthinspace~120 & 323 
$\pm$\negthinspace~11 $\pm$ 162 \\
Other backgrounds
                 &    43 $\pm$  ~4 $\pm$ ~13 &  51 $\pm$ ~4 $\pm$ 
\thinspace~15 \\
\hline
Total background &  2156 $\pm$ ~28 $\pm$\negthinspace~215 &1917 $\pm$ 
~26 $\pm$ 233 \\
\hline
Background subtracted
                 & 16775 $\pm$\negthinspace~183 $\pm$\negthinspace~215 
&8859 $\pm$\negthinspace~148 $\pm$ 233
$\pm$\negthinspace~148$\pm$ 233
\end{tabular}
\end{table}

\begin{table}[htbp]
\caption{ Goodness-of-fit for various models fitted to the primary 
spectrum.  Models modified to
vary the fraction of $D^{**}$ are indicated by '**', and the ``best
fit'' is described in the text.
}
\label{tab:chisq}
\begin{center}
\begin{tabular}{lc}
   Model & $\chi^2/DOF$ \\ \hline
   QQ98\cite{QQ} & 50.0/43 \\
   ISGW2\cite{ISGW} & 70.0/43 \\
   ISGW & 48.3/43 \\
   ACCMM\cite{ACCMM} & 125.8/43 \\
   ISGW2** & 32.5/41 \\
   QQ98 ** & 30.2/42\\
   QQ98 (best fit) & 28.4/39
\end{tabular}
\end{center}
\end{table}

\begin{table}[htbp]
\caption{ Summary of systematic errors on the branching fractions.
}
\label{sys_table}
\begin{center}
\begin{tabular}{lcc}
    Source of uncertainty                 &
    ${\Delta\cal{B}\over\cal{B}}$($B\rightarrow Xe^+\nu$)  (\%)&
    ${\Delta\cal{B}\over\cal{B}}$($B\rightarrow \bar{D}X,\bar{D}
\rightarrow Ye^-\bar{\nu}$)  (\%)\\
\hline \hline
Tracking                &   2.9 &   2.9  \\
Electron identification &  2.0  &   2.0  \\
Kinematic selection   &   1.4 &   4.1  \\
Continuum subtraction   &   0.7 &   0.9  \\
Background subtraction  &   1.8 &   2.3  \\
Mixing parameter        &   0.6 &   1.5  \\
Relative efficiency     &   1.3 &   0.8  \\
Model prediction        &   0.7 &  10.4  \\ \hline
Total                  &   4.5 &  12.1
\end{tabular}
\end{center}
\end{table}

\begin{figure}[htbp]
\centerline{\epsfysize=160mm \epsfbox{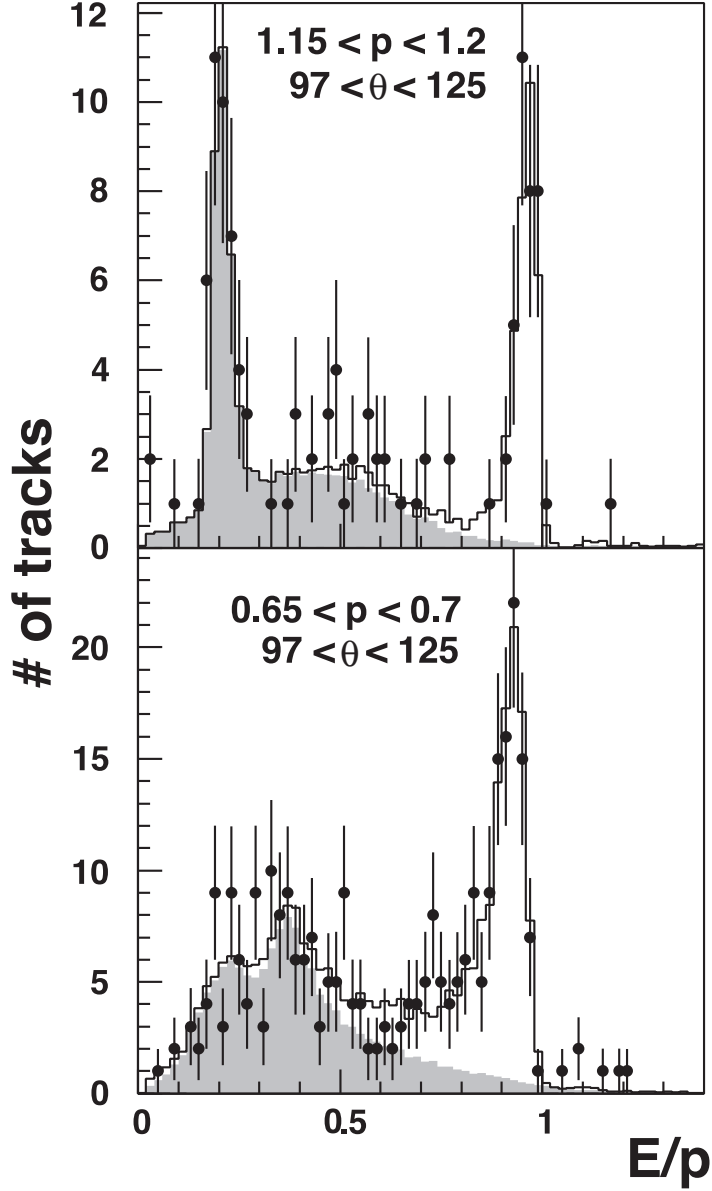}}
\caption{ $E/p$ distributions for electron candidates
in the momentum ranges 1.15--1.2~GeV/$c$ (top)
and 0.65--0.7~GeV/$c$ (bottom) with the electron candidate track
in the polar angle range between $97^\circ$ and $125^\circ$ in the
laboratury frame.
The points with error bars are the data,
the solid histogram  is the sum of signal and background components,
and the shaded  area shows the non-electron
background contribution.
   }
\label{eoverpfit}
\end{figure}

\begin{figure}[htbp]
\centerline{\epsfysize=150mm \epsfbox{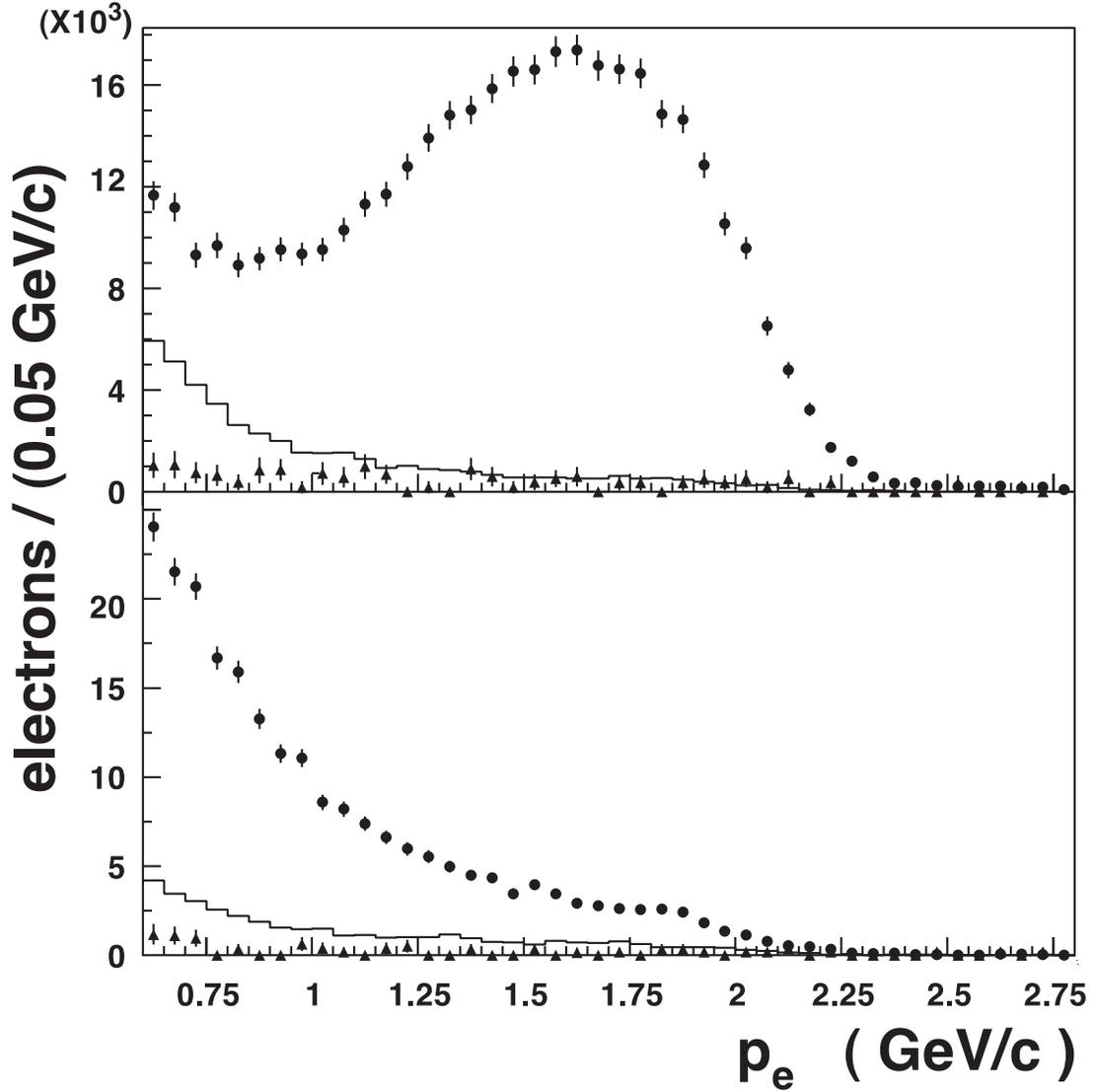}}
\caption{  Electron momentum spectra. The top and bottom figures show
the spectra of opposite-sign and same-sign electrons, respectively.
The closed circles are the on-resonance data.
The triangles show the scaled off-resonance data.
The error bars indicate only the statistical error.
The histogram is the MC-determined $B\bar{B}$ background.
}
\label{spectrum1}
\end{figure}

\begin{figure}
\centering
\mbox{\psfig{figure=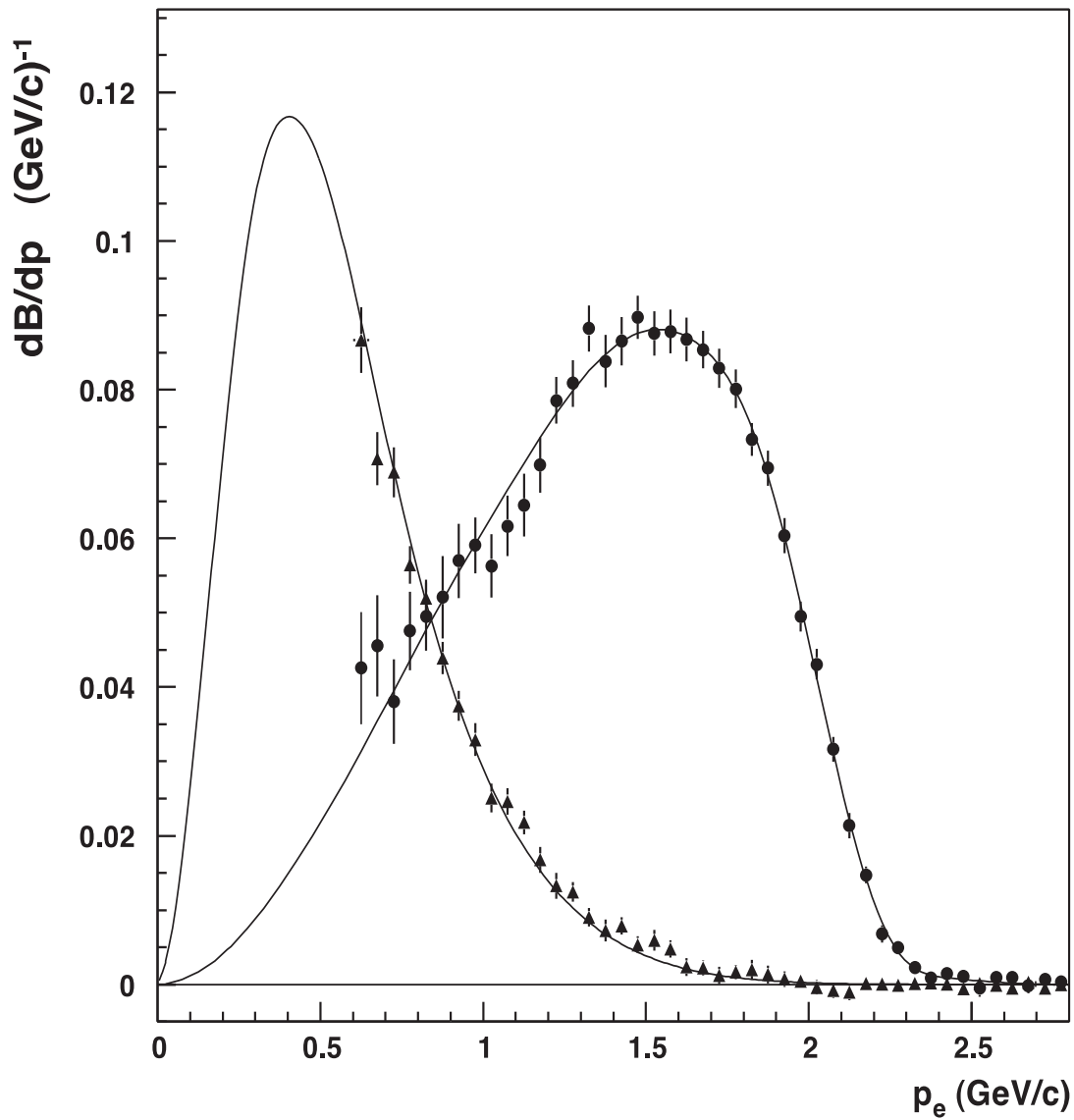,width=6.0in,height=6.0in}}
\centering
\caption{ Final results for the electron spectra of primary
(circles)
and secondary (triangles) semileptonic decays.
   The solid curve superimposed on the primary electron spectrum
is the best-fit model described in the text,
   while the curve for the secondary electron spectrum is the ISGW2 model.}
\label{final}
\end{figure}



\newpage

\begin{thebibliography}{99}
\bibitem{puzzle} I. Bigi, B. Blok, M. Shifman, A. Vainshtein,
                  Phys. Lett. {\bf B323} (1994) 408.
\bibitem{expA-br}  ARGUS Collaboration, H. Albrecht {\it et al.},
                  Phys. Lett.{\bf B318} (1993) 397.
\bibitem{expC-br}  CLEO Collaboration, B. Barish {\it et al.},
                  Phys. Rev. Lett. {\bf 76} (1997) 1570.
\bibitem{expZ-br} ALEPH Collaboration, D. Buskulic {\it et al.},
                  Z. Phys.{\bf C62} (1994) 179; \\
                   DELPHI Collaboration, P. Abreu {\it et al.},
                  Z. Phys. {\bf C66} (1995) 323; \\
                   L3 Collaboration, M. Acciarri {\it et al.},
                  Z. Phys. {\bf C71} (1996) 379; \\
                   OPAL Collaboration, G. Abbiendi {\it et al.},
                  Eur. Phys. J. {\bf C13} (2000) 225; \\
                   L3 Collaboration, M. Acciarri {\it et al.},
                  Eur. Phys. J. {\bf C13} (2000) 47;  \\
                  ALEPH, CDF, DELPHI, L3, OPAL, and SLD Collaborations, \\
                  D. Abbaneo {\it et al.},
                  SLAC-PUB-8492, CERN-EP-2000-096, \\
                  this paper indicates that
                  ${\cal B}(b\rightarrow xl\nu)$
                  is (10.58~$\pm$~0.07~$\pm$~0.17)\%.
\bibitem{th-ccs} A. F. Falk, M. B. Wise, I. Dunietz,
                  Phys. Rev. {\bf D51} (1995) 1183;\\
                  E. Bagan, Patricia Ball, V. M. Braun, P. Gosdzinsky,
                  Nucl. Phys. {\bf B432} (1994) 3;\\
                  E. Bagan, Patricia Ball, V. M. Braun, P. Gosdzinsky,
                  Phys. Lett. {\bf B342} (1995) 362;\\
                  erratum, Phys. Lett. {\bf B374} (1996) 363;\\
                  M. B. Voloshin,
                  Phys. Rev. {\bf D51} (1995) 3948.
\bibitem{b2ccs} M. Neubert and C. T. Sachrajda,
                 Nucl. Phys. {\bf B483} (1997) 339.
\bibitem{Nc_CLEO}{ CLEO Collaboration, L. Gibbons {\it et al.},
                  Phys. Rev. {\bf D56} (1997) 3783.
}
\bibitem{Nc_DELPHI}{ DELPHI Collaboration, P. Abreu {\it et al.},
                  Phys. Lett. {\bf B426} (1998) 193.
}
\bibitem{Nc_ALEPH}{ ALEPH Collaboration, D. Buskulic {\it et al.},
                  Phys. Lett. {\bf B388} (1996) 648.
}
\bibitem{KM} M. Kobayashi and T. Maskawa,
              Prog. Theor. Phys. {\bf 49} (1973) 652;\\
	     N. Cabibbo, Phys. Rev. Lett. {\bf 10} (1963) 531.
\bibitem{Belle}{BELLE Collaboration, A. Abashian {\it et al.},
		Nucl. Inst. and Meth. {\bf A479} (2002) 117.
   }
\bibitem{KEKB}{E Kikutani ed.,
		    KEK Preprint 2001-157 (2001),
		    to appear in Nucl. Instr. and Meth. A.}
\bibitem{MC}{
	R. Brun {\it et al.}, GEANT 3.21, CERN Report No. DD/EE/84-1 (1987).
}

\bibitem{FW}{
         G. Fox and S. Wolfram, Phys. Rev. Lett {\bf 41} (1978) 1581.
}
\bibitem{CLEO_DDX}{
         CLEO Collaboration, T. E.  Coan, {\it et al.},
         Phys. Rev. Lett {\bf 80} (1998) 1150.
}
\bibitem{PDG2000}  Particle Data Group, Eur. J. {\bf C15} (2000) 1.

\bibitem{ACCMM} G. Altarelli, N. Cabibbo, G. Corbo, L. Maiani,
  G. Martinelli,  Nucl. Phys. {\bf B208} (1982) 365.
\bibitem{ISGW} N. Isgur, D. Scora, B. Grinstein, M. B. Wise,
                Phys. Rev. {\bf D39} (1989) 799;\\
                D. Scora, N. Isgur,
                Phys. Rev. {\bf D52} (1995) 2783.
\bibitem{QQ} `QQ - The CLEO Event Generator',\\
     http://www.lns.cornell.edu/public/CLEO/soft/QQ (unpublished).
\bibitem{Bigi} I. Bigi, preprint hep-ph/9907270; \\
             I. Bigi, M. Shifman, N. Uraltsev,
             Ann. Rev. Nucl. Part. Sci. {\bf 47} (1997) 591.



\end{thebibliography}
\end{document}